# Visualizing acoustic levitation using COMSOL Multiphysics

Francisco M. Muñoz-Pérez∗, Juan C. Castro-Palacio,
Marcos H. Giménez and Juan A. Monsoriu

**Keywords:** Acoustic; levitation; transducer; COMSOL

**Abstract:** We present a new virtual laboratory developed with COMSOL Multiphysics for the simulation of an acoustic levitator. Our computer application simulates the acoustic pressure field and its interaction with a set of particles. Students can interact with the system by having the possibility of changing the frequency and distance parameters between transducers in real-time. We have also developed and shared for free use the 3D printing design files for the construction of necessary components for the acoustic levitator, as well as the instructions for its experimental implementation. The experimental results, along with the virtual laboratory, provide the students with useful tools to understand and interpret the acoustic phenomenon involved.

## 1 Introduction

Acoustic levitation has become a very attractive technology due to its diverse features [1]. One of them is that it allows the suspension of particles regardless of their composition as well as a manipulation environment in microgravity. Another relevant feature of acoustic levitation is that it allows the confinement of larger particles compared to light trapping techniques. Levitation has multiple applications such as the study of materials without being contained or requiring stable and non-invasive handling [2, 3]. Currently, this technology can be found in the study of small animals or liquids, or in the industrial sector in micro-assembly systems just to mention a few examples [4–6]. This type of levitator is based on a standing wave generated by ultrasonic transducers which produces a pattern for the acoustic pressure field. At the nodes of the pressure gradient, the particles present a dynamic equilibrium of forces, and it is at these points where the particles can be suspended [7, 8].

Students' exposure to this technology type allows them to approach acoustics research topics and their applications early. In addition, it reinforces the concepts of acoustic waves included in physics subjects for engineering. Previous work presents the study of acoustic levitators by addressing their construction [2, 9, 10]. However, a more didactic approach allows students to have a better understanding of this physical phenomenon. We believe that the combination of a virtual laboratory with an experimental system provides students with the means to develop their own understanding of the processes they are learning [11, 12]. Moreover, virtual labs give users the possibility of controlling different variables involved, so they can analyze their influence on the results. Within this context, virtual labs can strengthen the concepts of acoustics studied in the classroom and boost the motivation of the students.

---

[a]**Centro de Tecnologías Físicas, Universitat Politècnica de València, 46022 València, Spain.**
∗**Email: fmmuope1@upvnet.upv.es**

In this work, we present a new virtual laboratory developed with COMSOL multiphysics that allows numerical simulations for the visualization of the acoustic pressure field and the levitator nodes. COMSOL has emerged as software with multiple applications in science and technology due to its ease of use and flexibility [13, 14]. In addition, the design of an acoustic levitator for educational purposes and its low-cost experimental implementation with standard equipment is also included. The user interface allows the students to obtain the levitation positions by solving the physics problem numerically and also to make a comparison with what is observed experimentally.

## 2 Basic theory

Acoustic levitation is a particle suspension technique that seeks to compensate the effects of gravity through the interaction of particles with a pressure field of acoustic radiation. It consists in the generation of a stationary wave due to the acoustic radiation generated by an ultrasonic transducer and a reflector, or as in the case of this work by two transducers. If one reflector is placed, a large part of the acoustic wave is reflected and projected to the transducer. When two counter-propagating ultrasonic transducers are used on the same axis, an acoustic pressure field is generated in opposite directions forming a stationary acoustic radiation field. The superposition of these waves generates a distribution of maxima and minima (antinodes and nodes) of amplitude. Each ultrasonic transducer generates a wave of the same characteristics but in opposite directions,

$$P_1 = p_0 \, sin(kz - \omega t) \; and \; P_2 = p_0 \, sin(kz + \omega t), \tag{1}$$

where $p_0$ is the amplitude, $\lambda$ the wavelength, $k$ the wave number and $\omega$ the angular frequency. It has been chosen the z-axis as the direction of propagation of the wave. When $P_1$ and $P_2$ waves are superimposed, a standing wave of the sound field $P$ is generated, described by,

$$P = P_1 + P_2 = 2 \, p_0 \, cos(\omega t) \, sin(kz). \tag{2}$$

With this, the acoustic radiation pressure is obtained by means of $F = (5/6) \, \pi R^3 \, (\omega/\rho_f \, v^3) \, p^2 \, sin(2kz)$, donde $\rho_f$ is the air density and v is the sound velocity [15].

The described distance between the two sources must fulfill the condition of being an integer multiplied by half wavelength,

$$d = n\frac{\lambda}{2}. \tag{3}$$

As a particle interacts with the stationary acoustic field, it is ideally suspended at the nodes of the wave. At the nodes, the pressure tends to be minimal, causing a particle located there to experience a constant pressure that confines it to the center over the direction of propagation. By using the upper and lower pressure difference, it is possible to find a point of dynamic equilibrium where the sound radiation pressure acts as a restoring and counteracting force to the force of gravity. The position of the pressure nodes are spaced every $\frac{\lambda}{2}$, distributed along the radiation direction at $z = \frac{\lambda}{4}, 3\frac{\lambda}{4}, 5\frac{\lambda}{4}, \cdots$ (see Figure 1). Thus, the wavelength can be obtained through,

$$\lambda = \frac{v}{f}, \tag{4}$$

where $v$ is the speed of sound in the propagating medium (343 $m/s$ in air at 23°C) and $f$ the frequency of the acoustic wave.

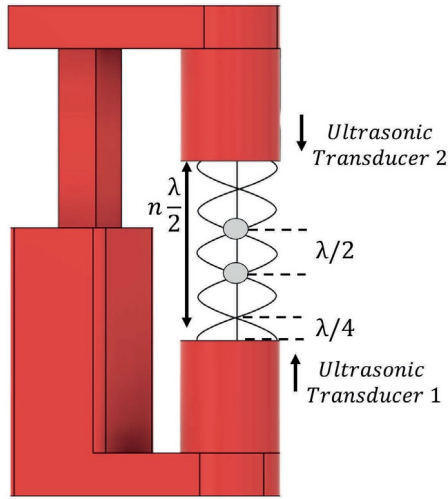

Figure 1: Schematic representation of acoustic levitation of counting-propagating transducers.

# 3 The program

A virtual laboratory has been developed with COMSOL Multiphysics 6.0 to simulate the acoustic radiation field originating from the proposed acoustic levitator [16]. Figure 2 shows the computation algorithm's flowchart consisting of three stages.

- The calculation process starts by choosing the input parameters and physical constants.
- A calculation section, which uses the parameters in the solution of the frequency domain inhomogeneous Helmholtz equation [17].
- Finally, the algorithm delivers the simulation of the acoustic pressure field.

Following this flowchart, we have implemented a COMSOL Multiphysics virtual lab whose main screen is shown in Figure 3. The interface is structured in these parts: the input parameters screen on the top left, an instructions section on the bottom left, and a results section on the right.

## 3.1 Input parameters

Figure 3 shows the input parameters panel of the user interface. As can be seen, the virtual laboratory allows entering the frequency $f$ of the sound wave and the distance $d$ between the ultrasonic transducers It also allows to choose whether to simulate the acoustic pressure field generated by both transducers or independently by transducer 1 (lower) or transducer 2 (upper). Another option available is the incorporation of particle interaction within the acoustic radiation field into the simulation (see Figure 3). This allows the student to observe the levitation points and make a comparison with those observed experimentally. It is important to mention that this increases the computation time since the virtual laboratory performs a higher number of calculations. The examples shown in this manuscript were implemented on a PC with 8 GB RAM and an RYZEN 7 Processor. The average time is approximately 5 sec and with particle simulation of $\approx 14$ sec. The calculation time depends on the characteristics of the PC where the virtual laboratory is implemented.

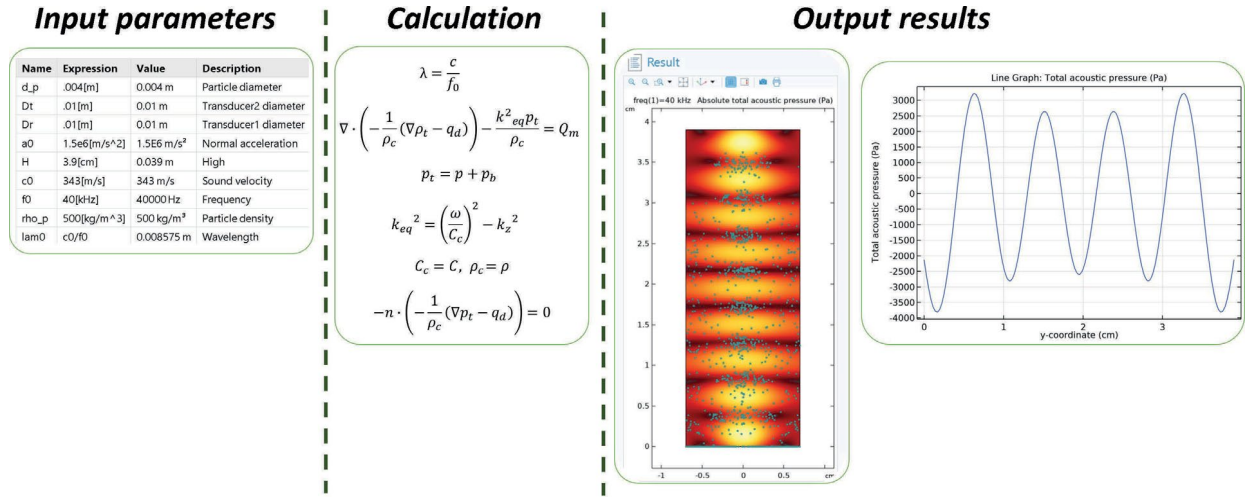

Figure 2: Flowchart of a virtual laboratory for acoustic levitation simulation.

### 3.2 Computation and results

Once the input parameters have been set, the user interface provides a "Compute" button to start the simulation process (see Figure 3). After a time-lapse, the result section shows the acoustic pressure field. If the option of simulation with particles has been activated previously, it will add particle interaction to the simulation result. Each time the input parameters are modified, it is necessary to click on the compute button to update the displayed results. For a better understanding of the results by the students, the virtual laboratory allows the generation of a more detailed report. By clicking the "Results", the results of the simulations are exported to a file in the folder of the user's choice. The report incorporates a summary, a table of parameters, an image of the absolute acoustic pressure field, and a plot of the total acoustic pressure intensity along the direction of propagation of both ultrasonic waves.

## 4 Results

The implementation of the acoustic levitator with polystyrene spheres has been carried out. The experimental setup shown in Figure 5 consists of two opposing ultrasonic transducers (PROWAVE, Mod.400SR160, $f = 40$ kHz), a 3D-printed holder that allows us to control the distance between transducers, and a frequency generator. The design of the 3D printed holder was carried out using Fusion 360 software. It was sought to be functional and to use a minimum of parts. It consists of two pieces; the lower one contains the first transducer and acts as a base. The second piece is the upper part, where the second transducer is placed. The holder has also an extendable part that allows to regulate the separation between both transducers. The 3D design is available at Thingsverse for free use [18]. The ultrasonic transducers are connected in parallel to the frequency generator. The latter is configured with a sinusoidal signal of $f = 40$ kHz and $Vpp = 10$ V. As shown in Figure 5, polystyrene spheres are placed at each node, seeking to levitate a particle at each node.

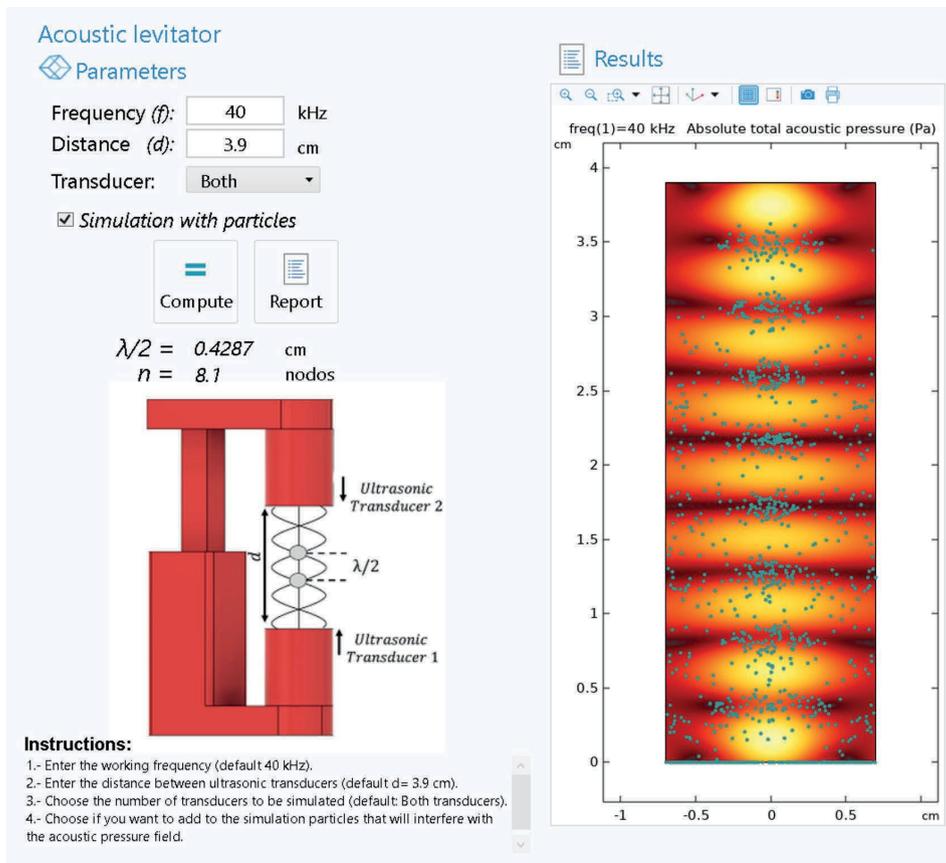

Figure 3: Virtual laboratory for acoustic levitation simulation.

As explained in the program section, COMSOL virtual labs allow us to observe the absolute acoustic pressure field for different distances between transducers. Figure 5 shows a set of results obtained through the virtual lab and particle levitation with the acoustic levitator for three different distances $d$: a) 2.0 cm, b) 2.5 cm, and c) 4.5 cm. For these cases, the frequency value used in the virtual laboratory is $f = 40\ kHz$. This is because it is the operating frequency of the transducers used in the acoustic levitator. As can be seen, as the distance increases, so does the number of nodes, in turn allowing us to have a larger number of suspended particles. The distance between each node maintains the relationship $\lambda/2$. This relationship can be seen in Figure 5.

As shown in Figure 5a), we performed the levitation of four polystyrene spheres and the absolute acoustic pressure field. Measuring the distance between the centers of two consecutive particles, we can observe a distance of 4.36 mm, while through the virtual laboratory, it is 4.4 mm. Both distances are very close to each other (4.36 mm ≈ 4.4 mm). Using the experimental distance obtained, $\lambda/2$ = 4.36 mm, we obtain the speed of sound: $v \approx$ 8.72 mm $x$ 40 kHz = 348.8 m/s. These values are very close to those reported in the literature for the speed of sound in air at 30°C [20], $v_a$ = 349.1 m/s, the same ambient temperature at the time of this experiment. The discrepancy percentage is less them 0.01%.

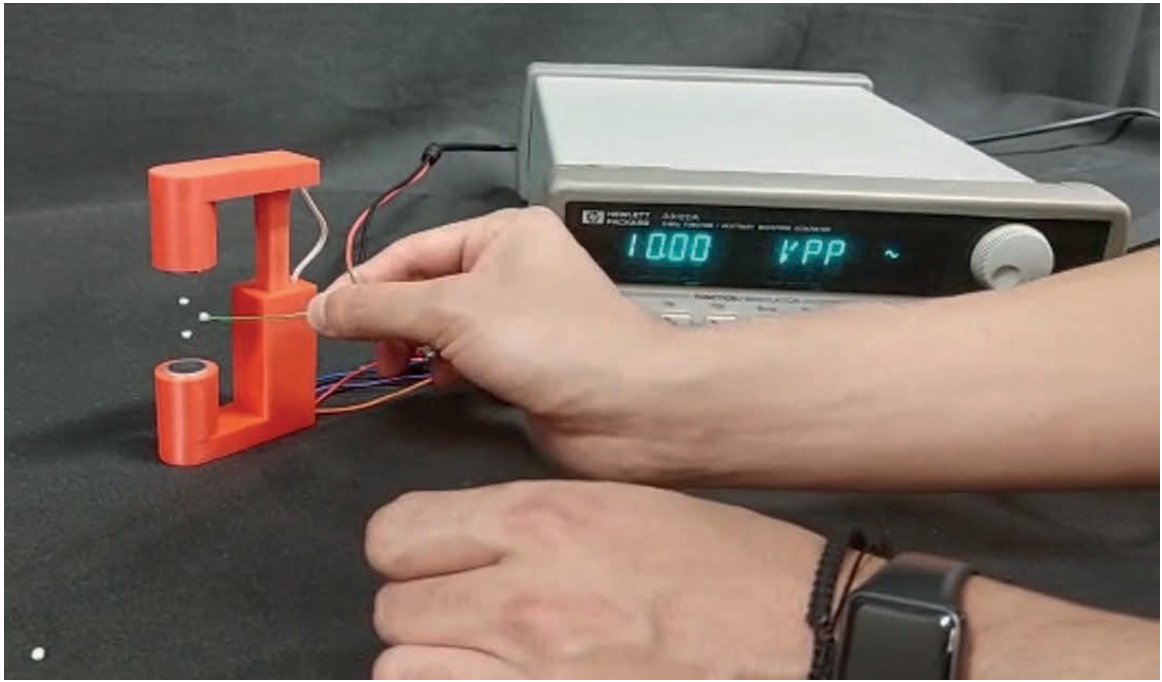

Figure 4: Experimental setup of acoustic levitator [[19]].

Another characteristic that we can check is the number of nodes generated in the standing wave. As seen in Figure 5 a-c, a comparison was made between the results obtained with the virtual laboratory and the experimental acoustic levitator. For a $d$ = 2 cm, the expected number of nodes is 4.66, as shown in Figure a) the levitation of 4 particles is possible. For 2 cm and 3 cm distances, the number of nodes is 5.83 and 10.49, respectively. As we can observe, in Figure 5 b-c, the number of levitated particles is 5 and 10, respectively.

## 5   Conclusions

Acoustic levitation is a physical phenomenon whose applications are expanding fast. This is why it is relevant for engineering students to know about it. A theoretical immersion accompanied by a virtual laboratory allows the students to understand more clearly the physical concepts of acoustic levitation. We provide for free use the files for the 3D printing of the levitator used in this work which is a low-cost alternative and easy to implement in a practical laboratory. The simulation through the virtual laboratory developed in COMSOL allows the students to corroborate the physical conditions and to visualize the acoustic pressure fields that are not appreciable experimentally.

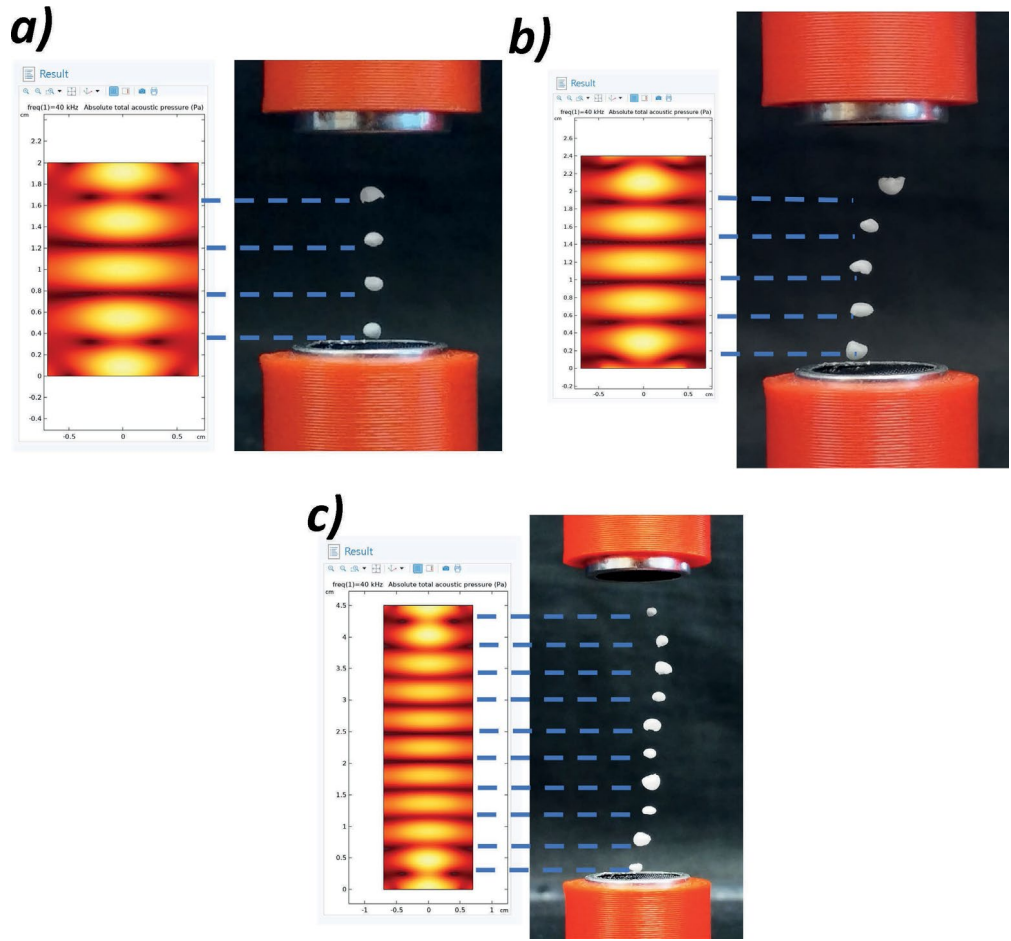

Figure 5: Results of the virtual laboratory and experimental for various distances between transducers for, a) $d$ = 2.0 cm, b) $d$ = 2.5 cm, and c) $d$ = 4.5 cm.

## References

[1] Al-Nuaimi, I. I. I., Mahyuddin, M. N. & Bachache, N. K. A non-contact manipulation for robotic applications: A review on acoustic levitation. *IEEE Access*, 10:120823–120837, 2022.

[2] Cordaro, R. & Cordaro, C. F. A demonstration of acoustical levitation. *The Physics Teacher*, 24(7):416, 1986.

[3] Morris, R. H., Dye, Elizabeth R., Docker, P. & Newton, M. I. Beyond the langevin horn: Transducer arrays for the acoustic levitation of liquid drops. *Physics of Fluids*, 31(10):101301, 2019.

[4] Watanabe, A., Hasegawa, K. & Abe, Y. Contactless fluid manipulation in air: Droplet coalescence and active mixing by acoustic levitation. *Scientific Reports*, 8(1):10221, 2018.

[5] Xie, W. J., Cao, C. D., Lü, Y. J., Hong, Z. Y. & Wei, B. Acoustic method for levitation of small living animals. *Applied Physics Letters*, 89(21):214102, 2006.

[6] Youssefi, O. & Diller, E. Contactless robotic micromanipulation in air using a magneto-acoustic system. *IEEE Robotics and Automation Letters*, 4(2):1580–1586, 2019.

[7] Crockett, A. & Rueckner, W. Visualizing sound waves with schlieren optics. *American Journal of Physics*, 86(11):870–876, 2018.

[8] Jackson, D. P. & Chang, Min-Hua. Acoustic levitation and the acoustic radiation force. *American Journal of Physics*, 89(4):383–392, 2021.

[9] Zhang, F. & Jin, Z. The experiment of acoustic levitation and the analysis by simulation. *Open Access Library Journal*, 5(10):1, 2018.

[10] Scott Schappe, R. & Barbosa, C. A simple, inexpensive acoustic levitation apparatus. *The Physics Teacher*, 55(1):6–7, 2017.

[11] Vidaurre, A., Riera, J., Giménez, M. H. & Monsoriu, J. A. Contribution of digital simulation in visualizing physics processes. *Computer Applications in Engineering Education*, 10(1):45–49, 2002.

[12] Daineko, Y., Dmitriyev, V., & Ipalakova, M. Using virtual laboratories in teaching natural sciences: An example of physics courses in university. *Computer Applications in Engineering Education*, 25(1):39–47, 2017.

[13] Riva, N., Grilli, F., & Dutoit, B. Superconductors for power applications: an executable and web application to learn about resistive fault current limiters. *European Journal of Physics*, 42(4):045802, 2021.

[14] Zhou, X. L., & Wang, J. H. Improvement of students' understanding about the phenomena of groundwater pumping by using computer software. *Computer Applications in Engineering Education*, 26(5):1792–1803, 2018.

[15] Yosioka, K. & Kawasima, Y. Acoustic radiation pressure on a compressible sphere. *Acta Acustica United with Acustica*, 5(3):167–173, 1955.

[16] Muñoz Pérez, F.M. Acoustic levitator. https://github.com/fmmuope/Acoustic-levitator.git, 2023. Accessed: 2023-06-28.

[17] COMSOL©. Comsol documentation: Acoustics module users guide. https://doc.comsol.com/6.0/doc/com.comsol.help.aco/AcousticsModuleUsersGuide.pdf, 2023. Accessed: 2023-06-26.

[18] Muñoz Pérez, F.M. Acoustic levitator holder. https://www.thingiverse.com/thing:6098283/files, 2023. Accessed: 2023-06-28.

[19] Muñoz Pérez, F.M. Levitating Objects Using Sound. https://media.upv.es/#/portal/video/afc70990-15a8-11ee-8071-7db45731f06b, 2023. Accessed: 2023-06-28.

[20] The Engineering ToolBox. Air - speed of sound vs. temperature. https://www.engineeringtoolbox.com/air-speed-sound-d_603.html, 2003. Accessed: 2023-06-26.